\documentclass[a4paper,UKenglish]{lipics-v2019}
\pdfoutput=1

\usepackage[dvipsnames]{xcolor}

\usepackage{amsmath,amssymb,amsthm,graphicx,fixmath,tikz,latexsym,xspace}

\newtheorem{wh}[theorem]{Working Hypothesis}

\title{Automated clustering of video games into groups with distinctive names}

\titlerunning{Automated clustering of video games into groups with distinctive names} 

\author{Nicolas Grelier}{Focus Entertainment, France}{nicolas.grelier@focusent.com}{}{}
\author{Stéphane Kaufmann}{Focus Entertainment, France}{stephane.kaufmann@focusent.com}{}{}

\authorrunning{N. Grelier, S. Kaufmann} 

\Copyright{Nicolas Grelier, Stéphane Kaufmann} 

\ccsdesc[500]{Applied computing~Publishing}

\keywords{Steam tags, naming score, Cohen's h}

\category{} 

\relatedversion{} 

\supplement{}

\nolinenumbers 

\hideLIPIcs  

\EventEditors{John Q. Open and Joan R. Access}
\EventNoEds{2}
\EventLongTitle{42nd Conference on Very Important Topics (CVIT 2016)}
\EventShortTitle{CVIT 2016}
\EventAcronym{CVIT}
\EventYear{2016}
\EventDate{December 24--27, 2016}
\EventLocation{Little Whinging, United Kingdom}
\EventLogo{}
\SeriesVolume{42}
\ArticleNo{23}
\begin{document}

\maketitle

\begin{abstract}
When doing a study on a large number of video games, it may be difficult to cluster them into coherent groups to better study them. In this paper, we introduce a novel algorithm, that takes as input any set of games $S$ that are released on Steam and an integer $k$, and cluster $S$ into $k$ groups. Each group is then assigned a distinctive name in the form of a Steam tag. We believe our tool to be valuable for gaining deeper insights into the video game market. We show that our algorithm maximises an objective function that we introduce, the naming score, which assesses the quality of a clustering and how distinctive its name is.
\end{abstract}

\section{Introduction}

In this paper, we are interested in understanding better the video game market. Let us imagine a scenario where Alice, who could be a researcher, a game developer, or working in a publishing company, faces a large number of games. She might be looking for relevant games for an academic study, or trying to find games similar to the one she is currently developing, to serve as a benchmark. The set of games could be all video games, or could be more specific: for instance all \emph{Racing} games, or all \emph{Puzzle} games released since $2018$ that have at least $1000$ reviews on the video game platform Steam.

Obviously, even as an expert, Alice only played a few of the games in the large set. She might have never heard of most of the games, especially those that did not sell well. Still, she wants to better understand this set. Maybe there are some games that are actually not relevant for her purposes, that she could then dismiss. Maybe she can learn that these games, which she thought were all similar, can actually be split into several coherent groups of games. This information may help Alice to do market segmentation~\cite{dolnicar2002review, saunders1980cluster}, since knowing the types of games that exist within a segment helps in understanding better the types of players. In this paper, we provide an algorithm for automatically splitting the games into groups with distinctive name, whatever the set of games with which Alice started.

To the best of our knowledge, there do not exist clustering algorithms for this specific task. One issue with video games clustering is that there do not seem to be clear ways of splitting games into groups, such that each game would clearly belong to a unique group, even when using genres~\cite{arsenault2009video,heintz2015game}. This is what one hopes to do with classical clustering, but we believe that for video games, one has to relax this desired property that a game should be as different as possible from the games in the other clusters. Thus, we introduce a new clustering objective. Our objective is to find an algorithm that could split a set of games into groups, and assign a name to each group. All games within a group should relate to its name. Moreover, the name should be as distinctive as possible, so that it gives information about the group. When facing different options, meaning that the names of several clusters relate to one game, we are fine with any arbitrary choice: We do not require that a game $G$ put into cluster $C$ would have nothing to do with the name of cluster $C'$. We only require that the name of cluster $C$ applies to $G$. In other words, we aim at high intra-cluster similarity, but lowering the inter-cluster similarity matters significantly less to us (we still want the name of a cluster to be distinctive enough from the rest of the games). Note that we do not require cluster names to be genres.

We want an algorithm that could partition any set of games into groups with distinctive names, using an explainable method, and that would reflect how players think. In this paper, we rely on data from the video game platform Steam, obtained through SteamSpy's API. On Steam, everyone can assign tags to games. As of November 2023, there are $448$ tags assigned to the $66908$ games available on Steam. Among the tags, one can find genre tags like \emph{Adventure} or \emph{RTS}, and many other tags like \emph{2D}, \emph{Funny}, \emph{Cats} or \emph{Medieval}. SteamSpy provides a database with for each game the twenty most assigned tags to it, along with how many players assigned those tags. We use those values for the clustering algorithms we test in this paper, and our cluster names are tags.

\subsection{Related work}

To the best of our knowledge, this is the first study about the problem of clustering games into groups with distinctive names, where the name of a cluster should apply to all the games it contains, but allowing that a game might have been placed in another cluster. Previous works about clustering games include clustering by genres through survey~\cite{heintz2015game}, by game traits through survey~\cite{fang2009lexical}, by players behaviours~\cite{bauckhage2014clustering}, based on how algorithms play games~\cite{ashlock2017general,bontrager2016matching}, and based on characteristics for video  streaming~\cite{zadtootaghaj2018classification}. Note that a major difference between these works and ours is that we want a program that could then be applied to cluster any set of games, whereas previous works aim at clustering all video games for good.

\subsection{Our contributions}

We introduce in Section~\ref{sec:obj} a new function to assess the quality of a clustering of video games. To maximise this function, the clustering must first have high intra-cluster similarity, and with less importance have low inter-cluster similarity, such that it is possible to give a name to each cluster of games that is distinctive from the other games. We call this function the \emph{naming score} of a clustering.

We provide in Section~\ref{sec:naming_score} a clustering algorithm that has significantly larger naming scores on several testing sets than simply applying a K-means algorithm, with and without Principal Component Analysis (PCA). Section~\ref{sec:applications} presents the results of our clustering algorithm applied to the testing sets: We provide the names of each cluster along with the titles of the five most reviewed games of that cluster.

\section{The naming score of a clustering}
\label{sec:obj}

We are given a set of games $S$ that we want to partition into $k$ clusters. Each cluster should have a distinctive name. Our idea is that we should be able to assign a tag $t$ to each cluster $C$, such that $t$ is assigned to most games in $C$ and to few games in $S \setminus C$. 

It is important to note a distinction between our clustering objective, and what is generally desired when doing clustering of some data points. Usually, it is wanted that a point should belong well to its cluster, and should be different from the points in the other clusters. This leads for instance to the well-known silhouette method to determine the number of clusters~\cite{rousseeuw1987silhouettes}. However, in our case, it is not an issue that a game might have been put in another cluster. We just want that the name of its cluster does strongly relate to the game. This is why we introduce a new objective function.

\subsection{Intuitive definition}

Let us consider one cluster $C$ of a clustering of a set of games $S$. We have access to the tags that are assigned by players to these games. Our idea is to find the most over-represented tag $t$ among those, and to name $C$ as the cluster of games that have the tag $t$. However, it is a priori not clear how one can say whether a tag is over-represented.

For this purpose, we use Cohen's h~\cite{cohen2013statistical}: a measure of difference between two proportions, in this case the proportion of games in cluster $C$ that have the tag $t$ compared to the proportion of games in the set $S$ that have the tag $t$. The exact formula for computing Cohen's h is given below. The reason why we use Cohen's h instead of, for instance, a simple ratio between the two proportions is as follows: Imagine there is a tag $t$ that is assigned to $0.001 \%$ of the games in $S$, and to $0.1 \%$ of the games in $C$. The increase in proportion is a multiplicative factor of $100$, but still the proportion of games in $C$ that have $t$ is only $0.1 \%$. Thus, it would be inappropriate to name $C$ as ``the group of games that have the tag $t$''. Cohen's h solves this issue by only giving a large value when the new proportion is both larger than the previous one, and the new proportion is in itself large.

With Cohen's h, we have a method of assessing whether a tag is over-represented in a group of games $C$. Therefore, we name $C$ as the ``the group of games that have the tag $t$'', where $t$ is the tag with highest Cohen's h value. We define the naming score of $C$ as the Cohen's h value of this tag $t$. Finally, we define the naming score of a clustering of $S$ as the weighted average naming score over all clusters, where the weight of a cluster $C$ is $|C|/|S|$. We do weighted average, for otherwise one could obtain a clustering with an extremely high naming score by putting nearly all games in one cluster (which would have a naming score close to $0$) and only one game in each other cluster (with a naming score close to $\pi$, i.e. the highest value Cohen's h may take). However, we want clusters to be roughly of the same size. By taking a weighted average, our naming score function may be large only if there is no large cluster. We give below a formal definition of this naming score.

\subsection{Formal definition}

Let us consider a set $S$ of video games and an integer $k$. The problem consists in partitioning $S$ into $k$ subsets, such that for each subset $C$ in $S$, there exists a tag $t$ that is distinctive to $C$. By ``distinctive'', we mean that $t$ is assigned to almost all games in $C$, and to few games in $S \setminus C$.

Let $C$ be a subset of video games in $S$ (potentially $S$ itself). We denote by $T_C$ the set of tags that are assigned to at least one game in $C$. For a tag $t$ in $T_C$, we denote by $p(t,C)$ the proportion of games in $C$ to which $t$ is assigned. We denote by $h(t,C)$ Cohen's h applied to the proportions $p(t,C)$ and $p(t,S)$, i.e. $h(t,C)$ is a measure of difference between the proportion of games in $C$ that have the tag $t$, and the proportion of games in $S$ that have the tag $t$. It is defined as $h(t,C) := \phi (p(t,C)) - \phi (p(t,S))$, where $\phi(x)$ equals $2 \arcsin{\sqrt{x}}$~\cite{cohen2013statistical}. Observe that the maximal value of Cohen's h is $\pi$. A greater Cohen's h value indicates a greater over-representation of the tag $t$ among the games in $C$ compared to the games in $S$.

We define the naming score $n(C)$ of a subset $C$ of $S$ as $n(C) := \max_{t \in T_C} h(t,C)$. Let $\mathcal{C} = \{C_i\}_{1\leq i \leq k}$ be a partition of $S$. We make an abuse of notation and denote by $n(\mathcal{C})$ the naming score of $\mathcal{C}$, which we define as $n(\mathcal{C}) := \frac{1}{|S|} \sum_{1\leq i \leq k} n(C_i)\cdot |C_i|$.

\section{Finding the best clustering algorithm}
\label{sec:naming_score}

In this paper, we use the classical K-means algorithm form the scikit-learn Python library to do the clusterings. We show that a modification of the Steam database (selecting only a few tags, weighting tags differently) significantly improves the naming score. We show step by step how we were able to improve the naming score by doing one modification after another. To speed up the running time of the K-means program, we only consider the $4065$ games that have at least $2000$ reviews on Steam.

\subsection{Priority helps}

In~\cite{grelier23icec}, we introduced the notion of priority of a steam tag. For a game $G$, where the most assigned tag $t_{\max}$ was assigned by $n_{\max}$ players, the priority of a tag $t$ that was assigned to $G$ by $n$ players is defined as $n/n_{\max}$. If $t$ is not assigned to $G$, then its priority is $0$. We argued that the priority is a measure of essentiality of tags to describe games. For a game $G$, tags with high priority are essential to describe $G$, and tags with low priority give information of less importance.

In this subsection, we show that priority helps to cluster games into groups with distinctive names. We compare two databases. In both, there is one entry per game, and as many columns as there are Steam tags. In the first database $\mathcal{D}_1$, there is a $1$ in the column of the tag $t$ for the game $G$ if the tag $t$ is assigned to $G$, and a $0$ otherwise. In the second database $\mathcal{D}_2$, the value at column $t$ for entry $G$ is the priority of the tag $t$ with respect to the game $G$.

\begin{figure}
    \centering
    \includegraphics{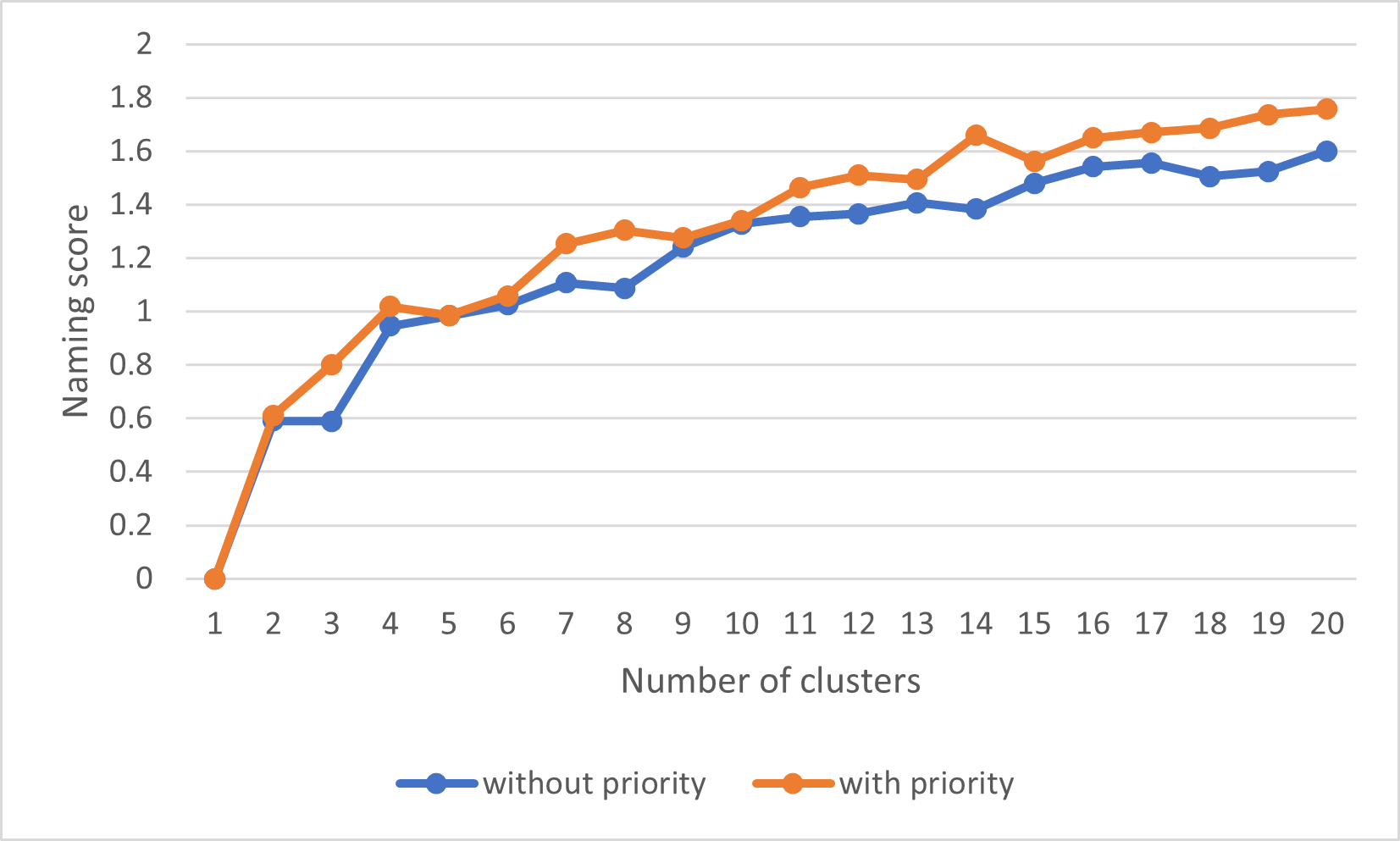}
    \caption{Comparison of the naming scores of the clusterings depending on whether the priority is taken into account.}
    \label{fig:no_prio}
\end{figure}

We compare the results of the K-means algorithm from the scikit-learn Python library applied to the two databases $\mathcal{D}_1$ and $\mathcal{D}_2$. Surprisingly, doing a Principal Component Analysis (PCA)~\cite{pearson1901liii} only lessened the naming scores. Thus, we apply the K-means algorithm directly to the databases for the rest of this paper. Figure~\ref{fig:no_prio} indicates that the naming scores of the clusterings are always larger using $\mathcal{D}_2$ than when using $\mathcal{D}_1$. Therefore, throughout the rest of the paper, we keep using the database $\mathcal{D}_2$.

\subsection{Restricting the set of tags}
\label{subsec:capital}

We have the intuition that the information contained in the Steam tags can be summarised using only a few tags. Our working hypothesis is as follows:

\begin{wh}\label{wh:weight_cap}
Having too many tags for the clustering brings noise. There exists a small set of tags that summarise all the important information contained in the Steam tags, which leads to better clusterings.
\end{wh}

This study was started in~\cite{grelier2023data}, where we presented a set of seven tags, the capital tags, that encompass all others. Those tags are \emph{Singleplayer}, \emph{Multiplayer}, \emph{Action}, \emph{Casual}, \emph{Adventure}, \emph{Strategy} and \emph{Anime}. Those tags where obtained by considering the tags that were correlated with many other tags, and that are assigned the most frequently. In~\cite{grelier2023data}, we say that a group of tags $\mathcal{T}$ \emph{covers} a set of games $S$ if for each game $G$ in $S$ there is at least one tag in $\mathcal{T}$ assigned to $G$. It was observed that the capital tags cover $94\%$ of all the games on Steam.

We iterate this process, by first removing the capital tags from the database. We refer now to those capital tags as being of rank $1$. To obtain the capital tags of rank $2$, we did an exhaustive search to find how to cover as many games as possible while using only a few tags. The number of games was chosen empirically, when we deemed that adding a new tag would not increase significantly the number of games covered. After one iteration of this process, we obtained the capital tags of rank $2$: \emph{RPG}, \emph{2D}, \emph{3D}, \emph{Atmospheric}, \emph{Simulation}, \emph{Colorful} and \emph{Puzzle}. After a second iteration, we obtained the capital tags of rank $3$: \emph{Pixel Graphics}, \emph{Funny}, \emph{Story Rich}, \emph{Fantasy}, \emph{Arcade}, \emph{Relaxing}, \emph{Shooter}, \emph{Management}, \emph{Horror}, \emph{Sci-fi}, \emph{Platformer}, \emph{Co-op}, \emph{Third Person}, \emph{Open World}, \emph{Rogue-like}, \emph{Exploration} and \emph{Sports}. We observe that out of the $4065$ games with at least $2000$ reviews, $98 \%$ have a capital tag of rank $1$, $96 \%$ have a capital tag of rank $2$, and $97 \%$ have a capital tag of rank $3$. Moreover, $99.7 \%$ have at least one capital tag (of any rank).

\begin{figure}
    \centering
    \includegraphics{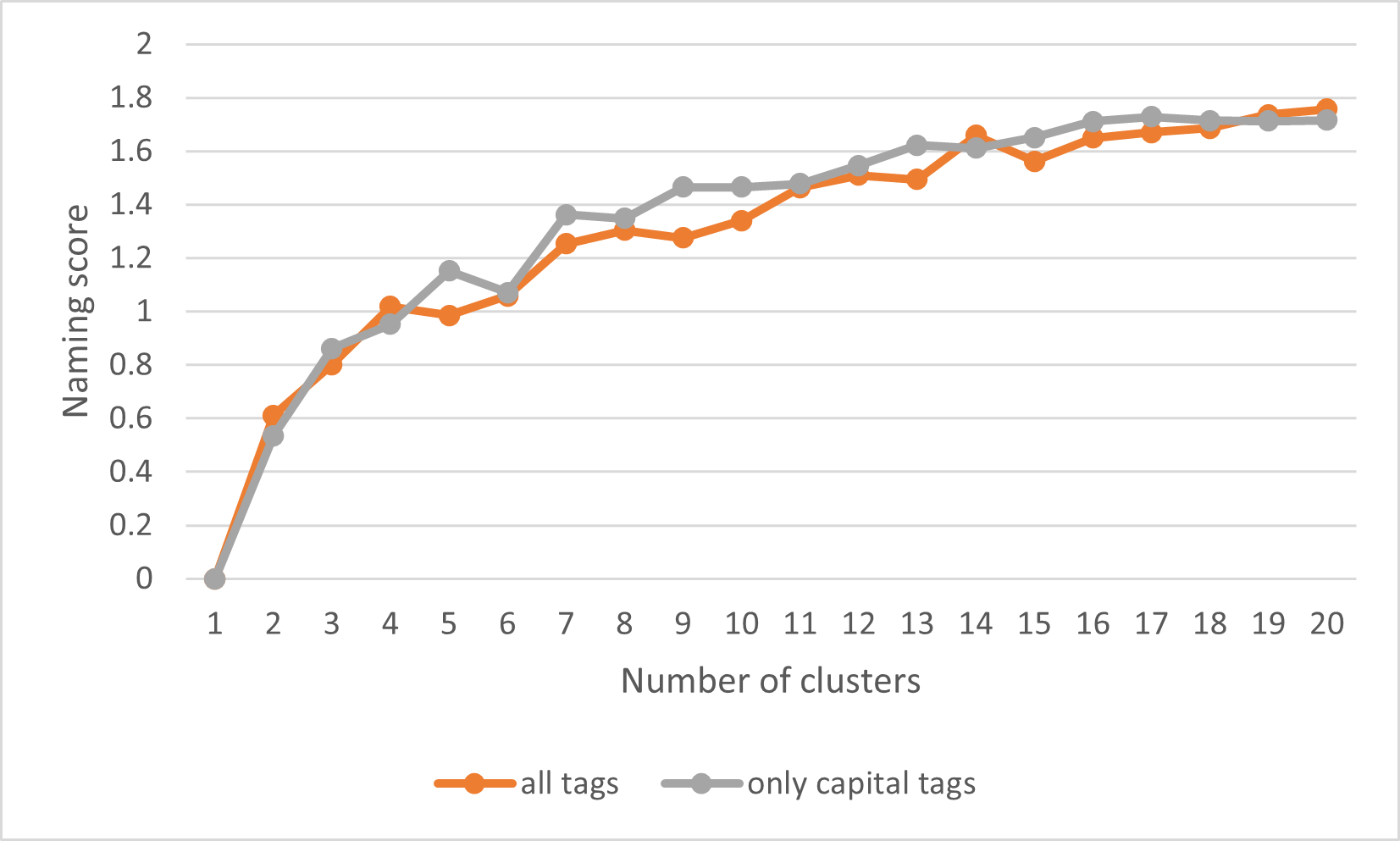}
    \caption{Comparison of the naming scores of the clusterings depending on whether they are computed on all tags or only on the capital tags.}
    \label{fig:capital}
\end{figure}

Figure~\ref{fig:capital} shows the naming scores when the K-means algorithm is done only on the capital tags of rank $1$, $2$ and $3$, compared to the previous results. Note that when we compute the naming score of a clustering, we do it on all Steam tags, not only on the capital tags. There are $14$ games out of the $4065$ that do not have any capital tags, and thus were not put into any cluster. Since we are missing only about $0.3 \%$ of the considered games, this difference is not significant. We observe that the naming scores are quite close, although they are generally better when using only the capital tags. 

\subsection{Using the ranks of the capital tags}

We have seen in Figure~\ref{fig:capital} that, for clustering the games into groups with distinctive names, using only the capital tags does not deteriorate the results. However, in Working Hypothesis~\ref{wh:weight_cap}, we stated the belief that capital tags do not only preserve the quality of the naming, but even improve it. We had the following intuition:

\begin{wh}
To obtain the best naming scores, capital tags should be treated differently according to their ranks.    
\end{wh}

Let us consider a game $G$. As said in Subsection~\ref{subsec:capital}, $G$ is very likely to have at least one capital tag of each rank. Let us consider only the capital tags of rank $i$ assigned to $G$. We tried several functions to apply to the priorities of the tags, we present here the one that gave the best naming scores. If $G$ has no capital tag of rank $i$, we do nothing. Otherwise, we can associate $G$ to a vector in $[0,1]^{d_i}$, where $d_i$ is the number of capital tags of rank $i$. The coordinate along axis $j$ is the priority of the capital tag $j$ of rank $i$ assigned to $G$. First, we normalise each vector by its $\ell_2$ norm. Secondly, we multiply it by a scaling factor denoted as $\lambda_i$, where $\lambda_1 = 0.25$, $\lambda_2 = 0.7$ and $\lambda_3 = 1$. Those values are the ones that maximised the naming scores in our experimentation.

\begin{figure}
    \centering
    \includegraphics{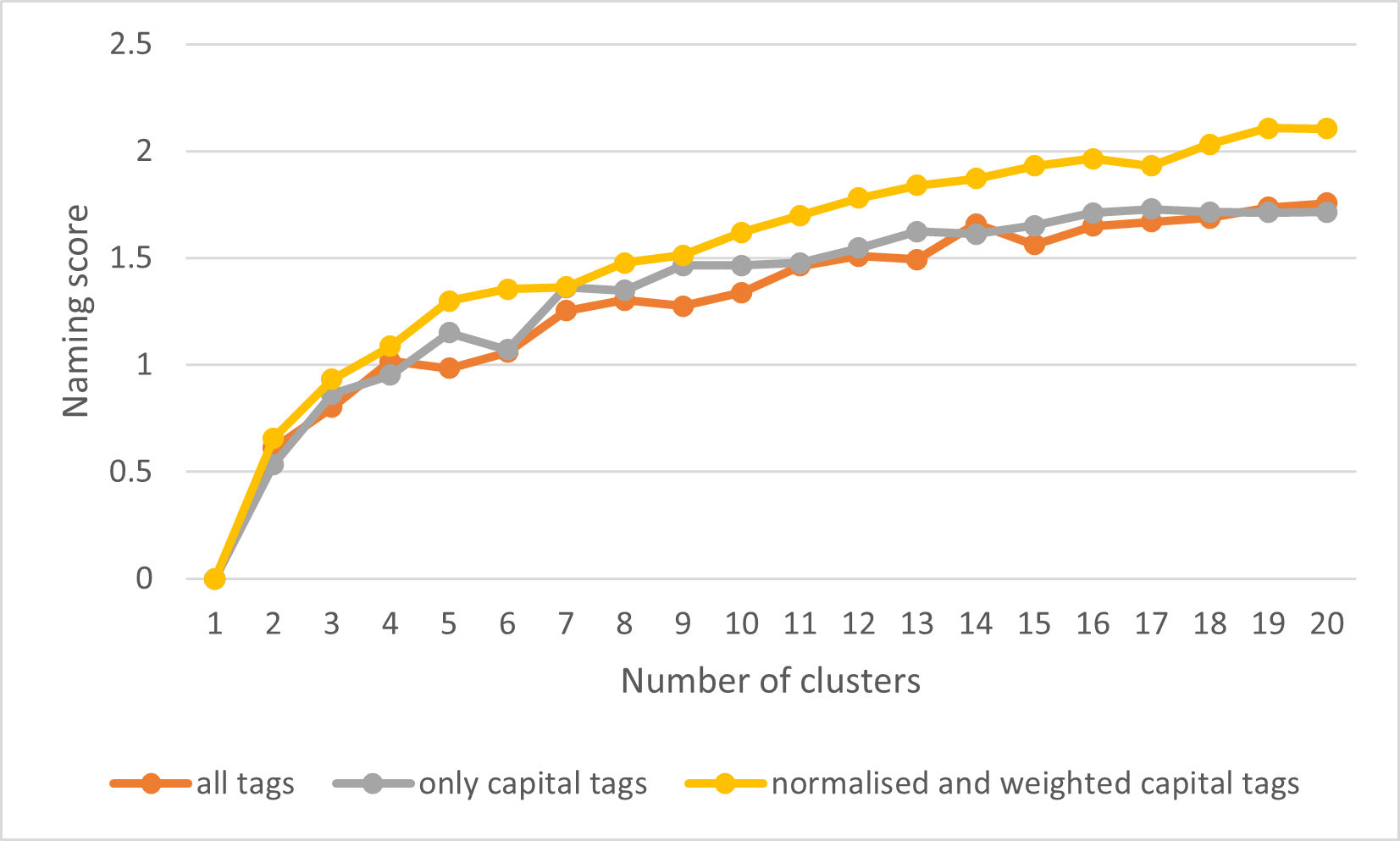}
    \caption{Comparison of the naming scores of the clusterings depending on whether they are computed on all tags, only on the capital tags or only on the normalised and weighted capital tags.}
    \label{fig:weight_capital}
\end{figure}

Figure~\ref{fig:weight_capital} shows a significant improvement of this new solution compared to the previous ones. This algorithm, using the database with priority and only keeping the normalised and weighted capital tags, is the algorithm that maximises the most the naming score among all the methods we tried for this set of games.

\subsection{Testing on other sets}

\begin{figure}
    \centering
    \includegraphics{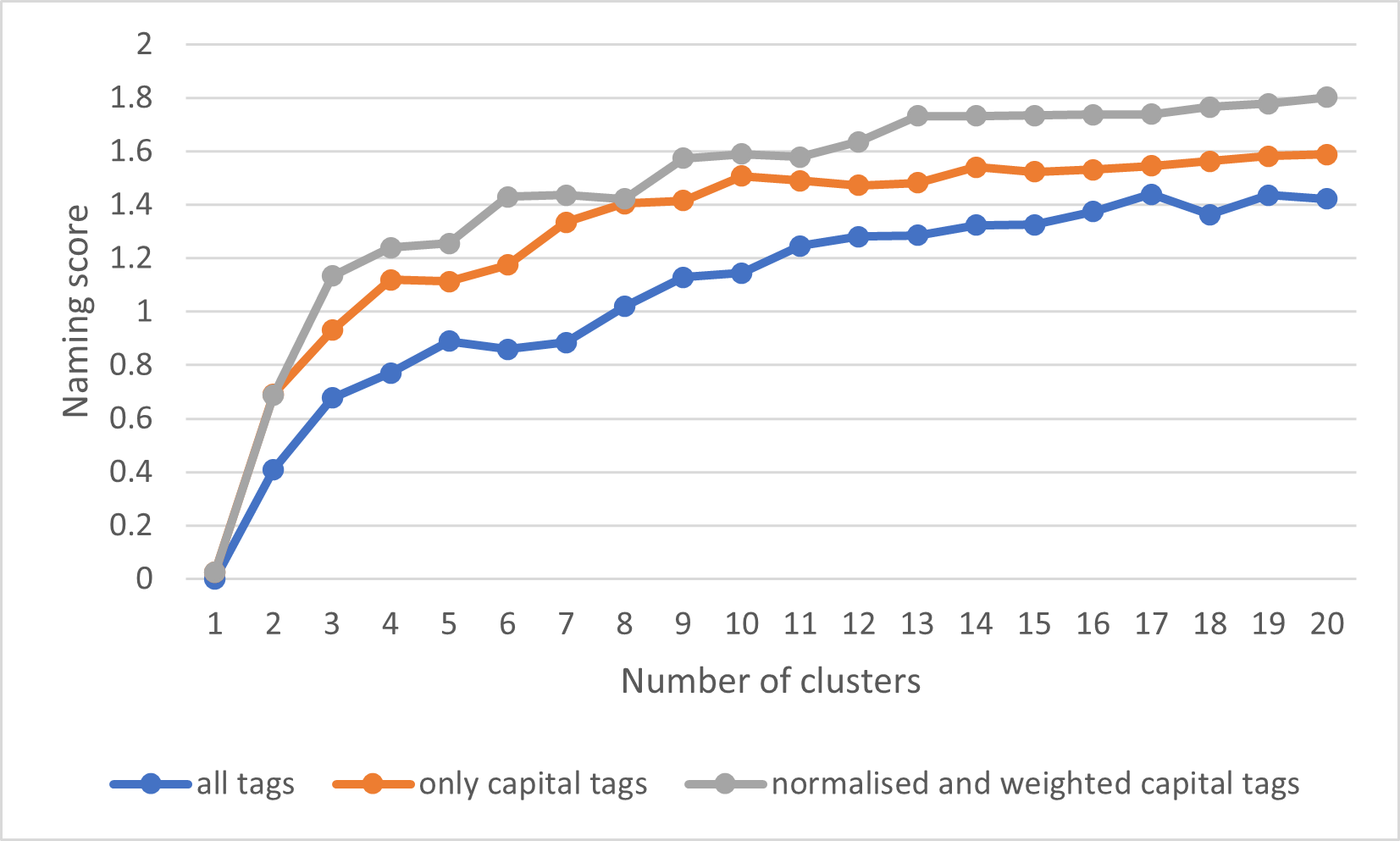}
    \caption{Comparisons of the three methods on the $2791$ games that have the tag \emph{Racing}.}
    \label{fig:racing}
\end{figure}
\begin{figure}
    \centering
    \includegraphics{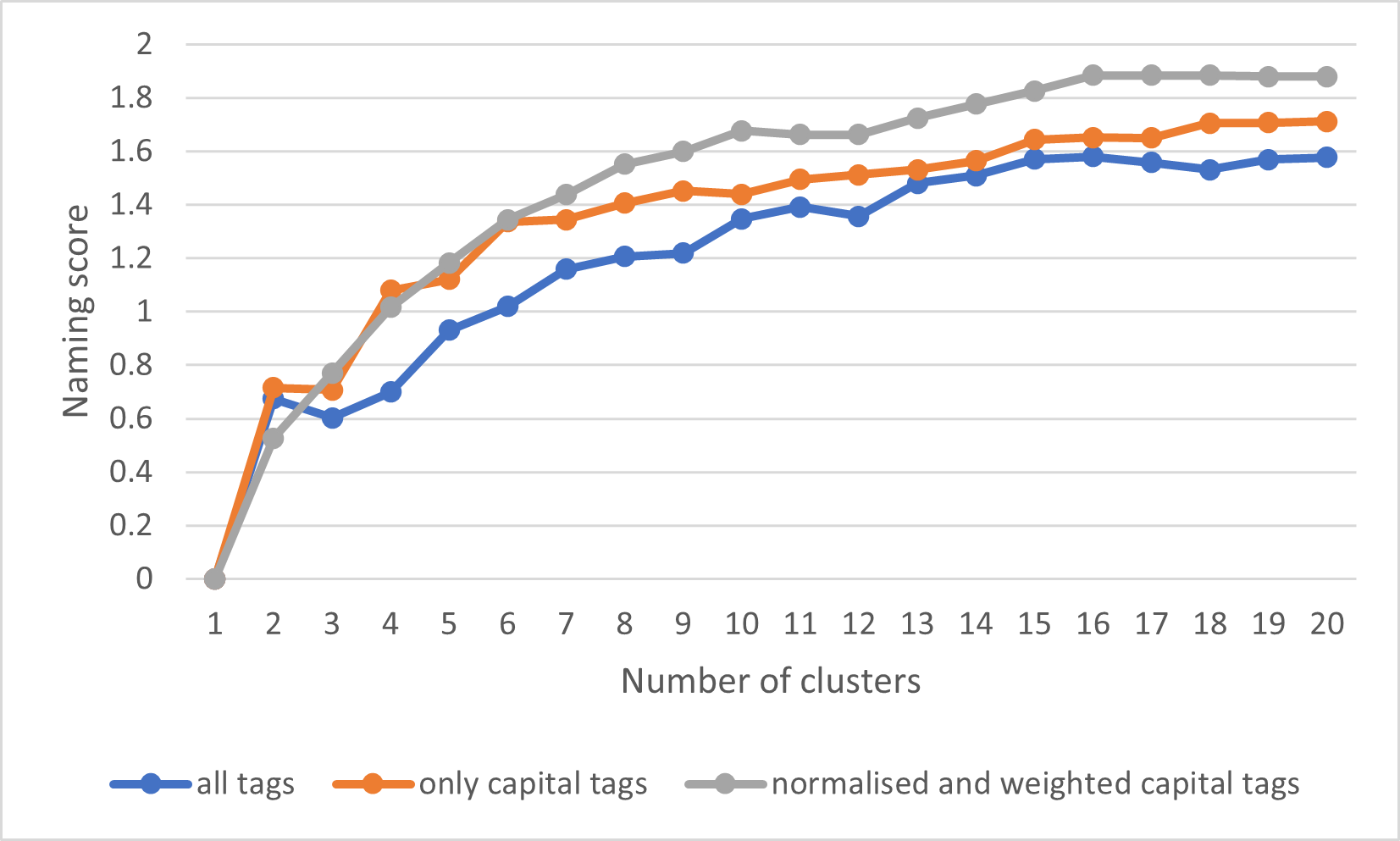}
    \caption{Comparisons of the three methods on the $12185$ games that have the tag \emph{Puzzle}.}
    \label{fig:puzzle}
\end{figure}

\begin{figure}
    \centering
    \includegraphics{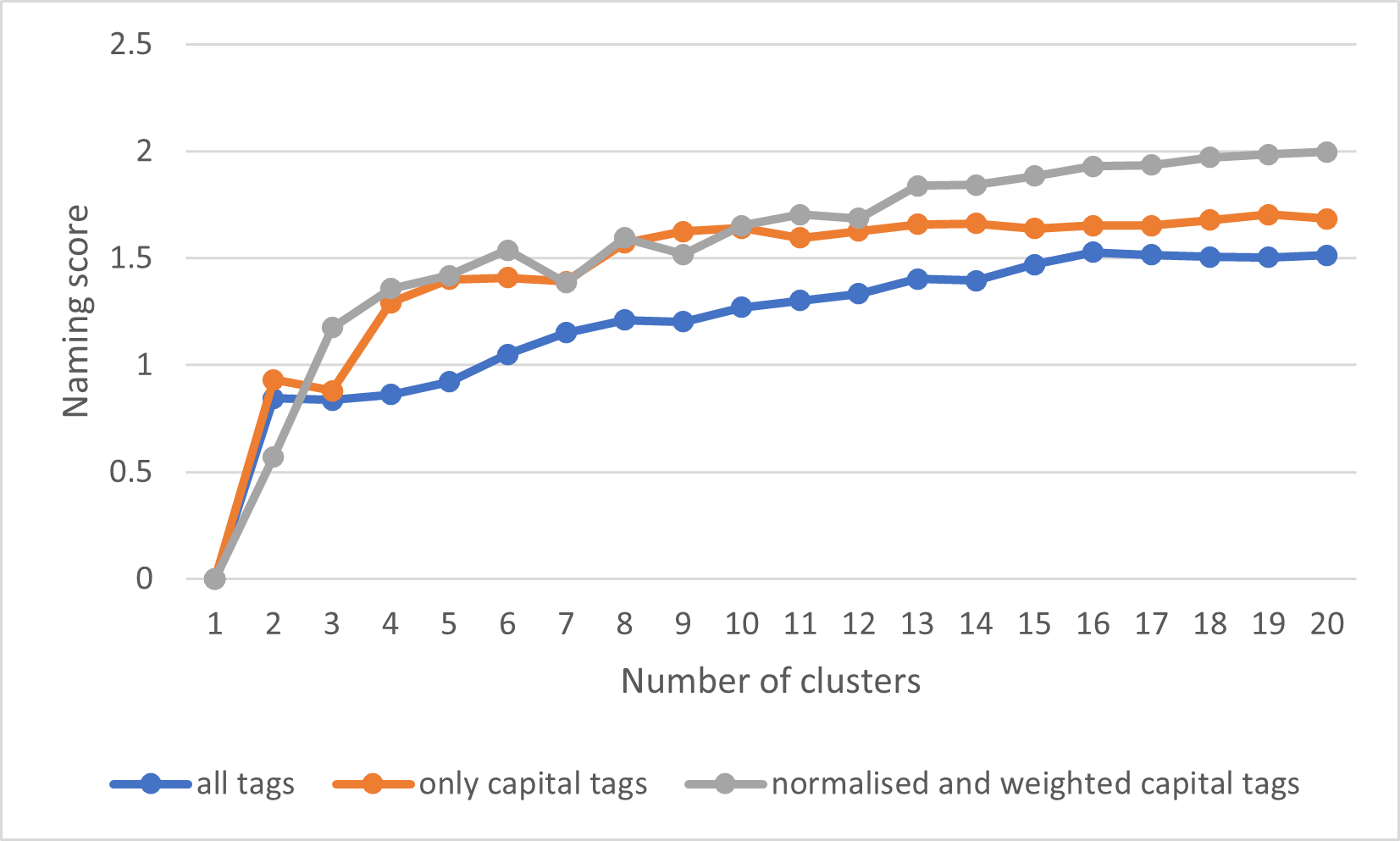}
    \caption{Comparisons of the three methods on the $2961$ games that have the tag \emph{Rogue-like}.}
    \label{fig:rogue-like}
\end{figure}

We tested the three methods of using the K-means algorithm, that is i) using all Steam tags, ii) using only the capital tags and iii) using the normalised and weighted capital tags, on other sets of games. Indeed, we want a program that would work well for all sets. We observe that Option iii) maximises the naming scores. Figures~\ref{fig:racing},~\ref{fig:puzzle} and~\ref{fig:rogue-like} depict the naming scores for the three methods on the sets of \emph{Racing} games, \emph{Puzzle} games and \emph{Rogue-like} games. We chose these sets of games as they are sufficiently large so that one would need a program and could not do a clustering by hand, not too large since we do not want the running times of the programs to be too long, and finally well-known because we provide in Section~\ref{sec:applications} the results of our best algorithm applied to these sets. As we want the reader to be able to compare our clusterings with their intuition, we needed set of games that are sufficiently well-known.

\section{Some applications}
\label{sec:applications}

In this section, we apply our best method, i.e. using the normalised and weighted capital tags, to some concrete examples, and present the cluster names. We also give for each cluster the titles of the five most reviewed games in that cluster (so that it is likely that the reader is familiar with them). As we have presented in Section~\ref{sec:naming_score} the curves of the naming scores with respect to the number of clusters, we can use the elbow method to choose an adequate number of clusters~\cite{ketchen1996application, kodinariya2013review, thorndike1953belongs}. The elbow method is not perfectly reliable, but it is sufficient for our purposes, since we just want to show that the cluster games and names are coherent.

For the clustering of the $4051$ games that have at least $2000$ reviews and at least one capital tag, we choose $8$ clusters. We give the list of clusters with their distinctive tag, along with their naming scores and the five most reviewed games in that clusters.

\begin{itemize}
    \item \emph{Sports} ($2.70$): Rocket League, EA SPORTS FIFA 23, FIFA 22, Assetto Corsa, TEKKEN~7.
    \item \emph{Anime} ($2.36$): Wallpaper Engine, NARAKA: BLADEPOINT, VRChat, Doki Doki Literature Club!, Mirror 2: Project X.
    \item \emph{Open World} ($1.98$): Grand Theft Auto V, Rust, ELDEN RING, Cyberpunk 2077, ARK: Survival Evolved.
    \item \emph{Simulation} ($1.67$): Euro Truck Simulator 2, War Thunder, Sid Meier's Civilization VI, Cities: Skylines, RimWorld.
    \item \emph{RPG} ($1.40$): The Witcher 3: Wild Hunt, Warframe, Dying Light, Baldur's Gate 3, The Elder Scrolls V: Skyrim.
    \item \emph{2D} ($1.14$): Terraria, Stardew Valley, Hollow Knight, Geometry Dash, The Binding of Isaac: Rebirth.
    \item \emph{Atmospheric} ($1.11$): Dead by Daylight, Portal 2, Sekiro: Shadows Die Twice, Deep Rock Galactic, Tomb Raider.
    \item \emph{Multiplayer} ($1.01$): Counter-Strike: Global Offensive, PUBG: BATTLEGROUNDS, Dota 2, Tom Clancy's Rainbow Six Siege, Garry's Mod.
\end{itemize}

There are no real surprises in those lists, except maybe for Wallpaper Engine and VRChat being in the \emph{Anime} cluster. This is not a mistake from the program, since \emph{Anime} has indeed a high priority in both of these games. Thus, this result truly corresponds to how the Steam community perceives these games. We use $6$ clusters for the $2091$ games with the tag \emph{Racing} that have at least one capital tag:

\begin{itemize}
    \item \emph{Open World} ($2.39$): Grand Theft Auto V, BeamNG.drive, Forza Horizon 4, Forza Horizon 5, Need for Speed Heat.
    \item \emph{Third Person} ($2.25$): Sonic Adventure 2, MX Bikes, Sonic Adventure DX, Road Redemption, Mon Bazou.
    \item \emph{Arcade} ($1.87$): TrackMania Nations Forever, FlatOut 2, Retrowave, Garfield Kart - Furious Racing, AudioSurf.
    \item \emph{Sports} ($1.85$): Rocket League, F1 2021, SpeedRunners, Need For Speed: Hot Pursuit, Sonic \& All-Stars Racing Transformed Collection.
    \item \emph{Simulation} ($1.73$): Assetto Corsa, CarX Drift Racing Online, F1 2020, DiRT Rally, DiRT Rally 2.0.
    \item \emph{Indie} ($0.23$): Stumble Guys, Crossout, Transformice, Notes of Soul, Car Mechanic Simulator 2015.
\end{itemize}

We have $8$ clusters for the $12185$ games with the tag \emph{Puzzle} (that necessarily have at least one capital tag since \emph{Puzzle} is a capital tag of rank $2$):
\begin{itemize}
    \item \emph{Platformer} ($2.42$): Ori and the Will of the Wisps, BattleBlock Theater, GRIS, Ori and the Blind Forest: Definitive Edition, Ori and the Blind Forest.
    \item \emph{Pixel Graphics} ($2.36$): Papers, Please, The Escapists 2, CARRION, There Is No Game: Wrong Dimension, Baba Is You.
    \item \emph{Horror} ($2.34$): Resident Evil 2, Inscryption, Little Nightmares, Resident Evil 7 Biohazard, DEVOUR.
    \item \emph{Exploration} ($2.19$): Stray, Outer Wilds, Inside the Backrooms, SOMA, Atomic Heart.
    \item \emph{Story Rich} ($2.17$): Portal 2, Tomb Raider, Half-Life 2, Portal, Batman: Arkham Knight.
    \item \emph{Relaxing} ($2.14$): PC Building Simulator, Dorfromantik, Unpacking, Mini Motorways, ISLANDERS.
    \item \emph{Adventure} ($0.52$): Hogwarts Legacy, STAR WARS Jedi: Fallen Order, Rise of the Tomb Raider, The Stanley Parable, Mirror's Edge.
    \item \emph{Casual} ($0.19$): Human Fall Flat, It Takes Two, Helltaker, Plants vs. Zombies GOTY Edition, Mirror.
\end{itemize}

We choose $6$ clusters for the $2961$ games with the tag \emph{Rogue-like} (that necessarily have at least one capital tag since \emph{Rogue-like} is a capital tag of rank $3$). We observe that there are two clusters that share the same tag with highest Cohen's h: \emph{Pixel Graphics}. This is not a contradiction to our objective, since we are fine with having games relating to more than one cluster. However, if one wants to better understand the differences between the two clusters, it is possible to consider for each cluster the two tags with highest Cohen's h, instead of only the highest one. In this case, one cluster would be named \emph{Pixel Graphics}-\emph{2D} and the other \emph{Pixel Graphics}-\emph{RPG}. We discuss this further in Section~\ref{sec:concl}.

\begin{itemize}
    \item \emph{Arcade} ($2.30$): Paint the Town Red, Crab Champions, Soulstone Survivors, Rogue Legacy 2, Nova Drift.
    \item \emph{Pixel Graphics}-\emph{RPG} ($1.83$): Noita, Realm of the Mad God Exalt, Hero Siege, Loop Hero, Stoneshard.
    \item \emph{Pixel Graphics}-\emph{2D} ($1.82$): The Binding of Isaac: Rebirth, Vampire Survivors, Dead Cells, Enter the Gungeon, Skul: The Hero Slayer.
    \item \emph{RPG} ($1.66$): Hades, Slay the Spire, Darkest Dungeon, The Scroll Of Taiwu, The Binding of Isaac.
    \item \emph{2D} ($1.52$): FTL: Faster Than Light, Rogue Legacy, Spelunky 2, Spelunky, Have a Nice Death.
    \item \emph{3D} ($0.60$): Risk of Rain 2, Don't Starve, Inscryption, Gunfire Reborn, Lobotomy Corporation | Monster Management Simulation.
\end{itemize}

There are issues with the last cluster, \emph{3D}, which contains games that do not use 3D game graphics. This can be partly understood by the fact that its naming score is quite low, $0.60$. Therefore, our clustering algorithm failed in this case to have all clusters with a distinctive name. However, it is unclear whether it is truly possible to do better when clustering the \emph{Rogue-like} games into $6$ clusters.

\section{Conclusion}
\label{sec:concl}

We have introduced a new optimisation function for clustering games into groups with distinctive names. We showed that using the priority notion on normalised and weighted capital tags improves the naming score of the clusterings when using the K-means algorithm. We also provided a few example of clusterings that are coherent. We noticed that two clusters may share the same tag with highest Cohen's h. It could be interesting not only to consider the tag with highest Cohen's h, but for instance the three best tags. The name of a cluster would then not consist in one tag but in the three tags with highest Cohen's h. We conjecture that our algorithm would still provide better naming scores than the other algorithms we tested in this paper, but we leave this question to future work. Our clustering algorithm works only on games that have at least one capital tag. As $99.7\%$ of the games with at least $2000$ reviews have at least one capital tag, this is not too problematic. Still, it would be interesting to try to guess for the remaining games what their capital tags could have been. We believe this could be done using traditional machine learning methods.




\newpage
\bibliography{bib}

\begin{thebibliography}{10}

\bibitem{arsenault2009video}
Dominic Arsenault.
\newblock Video game genre, evolution and innovation.
\newblock {\em Eludamos: Journal for computer game culture}, 3(2):149--176, 2009.

\bibitem{ashlock2017general}
Daniel Ashlock, Diego Perez-Liebana, and Amanda Saunders.
\newblock General video game playing escapes the no free lunch theorem.
\newblock In {\em 2017 IEEE Conference on Computational Intelligence and Games (CIG)}, pages 17--24. IEEE, 2017.

\bibitem{bauckhage2014clustering}
Christian Bauckhage, Anders Drachen, and Rafet Sifa.
\newblock Clustering game behavior data.
\newblock {\em IEEE Transactions on Computational Intelligence and AI in Games}, 7(3):266--278, 2014.

\bibitem{bontrager2016matching}
Philip Bontrager, Ahmed Khalifa, Andre Mendes, and Julian Togelius.
\newblock Matching games and algorithms for general video game playing.
\newblock In {\em Proceedings of the AAAI Conference on Artificial Intelligence and Interactive Digital Entertainment}, pages 122--128, 2016.

\bibitem{cohen2013statistical}
Jacob Cohen.
\newblock {\em Statistical power analysis for the behavioral sciences}.
\newblock Academic press, 2013.

\bibitem{dolnicar2002review}
Sara Dolnicar.
\newblock A review of unquestioned standards in using cluster analysis for data-driven market segmentation.
\newblock 2002.

\bibitem{fang2009lexical}
Xiaowen Fang, Susy~S Chan, and Chitra Nair.
\newblock A lexical approach to classifying computer games.
\newblock 2009.

\bibitem{grelier23icec}
Nicolas Grelier and St{\'{e}}phane Kaufmann.
\newblock A data-driven classification of video game vocabulary.
\newblock In {\em Entertainment Computing - {ICEC} 2023 - 22nd {IFIP} {TC} 14 International Conference}, volume 14455 of {\em Lecture Notes in Computer Science}, pages 17--30. Springer, 2023.
\newblock \href {https://doi.org/10.1007/978-981-99-8248-6\_2} {\path{doi:10.1007/978-981-99-8248-6\_2}}.

\bibitem{grelier2023data}
Nicolas Grelier and St{\'e}phane Kaufmann.
\newblock Data-driven classifications of video game vocabulary.
\newblock {\em arXiv preprint arXiv:2303.07179}, 2023.

\bibitem{heintz2015game}
Stephanie Heintz and Effie Lai-Chong Law.
\newblock The game genre map: A revised game classification.
\newblock In {\em Proceedings of the 2015 annual Symposium on computer-human Interaction in play}, pages 175--184, 2015.

\bibitem{ketchen1996application}
David~J Ketchen and Christopher~L Shook.
\newblock The application of cluster analysis in strategic management research: an analysis and critique.
\newblock {\em Strategic management journal}, 17(6):441--458, 1996.

\bibitem{kodinariya2013review}
Trupti~M Kodinariya and Prashant~R Makwana.
\newblock Review on determining number of cluster in {K}-means clustering.
\newblock {\em International Journal}, 1(6):90--95, 2013.

\bibitem{pearson1901liii}
Karl Pearson.
\newblock On lines and planes of closest fit to systems of points in space.
\newblock {\em The London, Edinburgh, and Dublin philosophical magazine and journal of science}, 2(11):559--572, 1901.

\bibitem{rousseeuw1987silhouettes}
Peter~J Rousseeuw.
\newblock Silhouettes: a graphical aid to the interpretation and validation of cluster analysis.
\newblock {\em Journal of computational and applied mathematics}, 20:53--65, 1987.

\bibitem{saunders1980cluster}
John~A Saunders.
\newblock Cluster analysis for market segmentation.
\newblock {\em European Journal of marketing}, 14(7):422--435, 1980.

\bibitem{thorndike1953belongs}
Robert~L Thorndike.
\newblock Who belongs in the family?
\newblock {\em Psychometrika}, 18(4):267--276, 1953.

\bibitem{zadtootaghaj2018classification}
Saman Zadtootaghaj, Steven Schmidt, Nabajeet Barman, Sebastian M{\"o}ller, and Maria~G Martini.
\newblock A classification of video games based on game characteristics linked to video coding complexity.
\newblock In {\em 2018 16th Annual workshop on network and systems support for games (NetGames)}, pages 1--6. IEEE, 2018.

\end{thebibliography}

\appendix
\section{Tables corresponding to the figures of the paper}

\begin{table}[ht]
\centering
\resizebox{\columnwidth}{!}{%
\begin{tabular}{ccccc}
\multicolumn{1}{l}{number of clusters} & without priority & all tags   & \multicolumn{1}{l}{capital tags} & \multicolumn{1}{l}{normalised and weighted capital tags} \\
1                                      & \textbf{0}       & \textbf{0} & \textbf{0}                       & \textbf{0}                                               \\
2                                      & 0.59             & 0.61       & 0.54                             & \textbf{0.66}                                            \\
3                                      & 0.59             & 0.8        & 0.86                             & \textbf{0.93}                                            \\
4                                      & 0.95             & 1.02       & 0.95                             & \textbf{1.09}                                            \\
5                                      & 0.98             & 0.98       & 1.15                             & \textbf{1.3}                                             \\
6                                      & 1.02             & 1.06       & 1.07                             & \textbf{1.35}                                            \\
7                                      & 1.11             & 1.25       & 1.36                             & \textbf{1.37}                                            \\
8                                      & 1.09             & 1.3        & 1.35                             & \textbf{1.48}                                            \\
9                                      & 1.24             & 1.28       & 1.47                             & \textbf{1.51}                                            \\
10                                     & 1.33             & 1.34       & 1.47                             & \textbf{1.62}                                            \\
11                                     & 1.36             & 1.46       & 1.48                             & \textbf{1.7}                                             \\
12                                     & 1.37             & 1.51       & 1.55                             & \textbf{1.78}                                            \\
13                                     & 1.41             & 1.49       & 1.62                             & \textbf{1.84}                                            \\
14                                     & 1.38             & 1.66       & 1.61                             & \textbf{1.87}                                            \\
15                                     & 1.48             & 1.56       & 1.65                             & \textbf{1.93}                                            \\
16                                     & 1.54             & 1.65       & 1.71                             & \textbf{1.96}                                            \\
17                                     & 1.56             & 1.67       & 1.73                             & \textbf{1.93}                                            \\
18                                     & 1.5              & 1.69       & 1.71                             & \textbf{2.03}                                            \\
19                                     & 1.52             & 1.74       & 1.71                             & \textbf{2.11}                                            \\
20                                     & 1.6              & 1.76       & 1.72                             & \textbf{2.1}                                            
\end{tabular}}
\caption{Naming scores of the algorithms presented in Section~\ref{sec:naming_score} applied to the $4065$ games with at least $2000$ reviews.}
\label{table:all_games}
\end{table}
\begin{table}[ht]
\centering
\begin{tabular}{cccc}
\multicolumn{1}{l}{number of clusters} & all tags   & \multicolumn{1}{l}{capital tags} & \multicolumn{1}{l}{normalised and weighted capital tags} \\
1                                      & \textbf{0} & \textbf{0}                       & \textbf{0}                                               \\
2                                      & 0.41       & \textbf{0.69}                    & \textbf{0.69}                                            \\
3                                      & 0.68       & 0.93                             & \textbf{1.13}                                            \\
4                                      & 0.77       & 1.12                             & \textbf{1.24}                                            \\
5                                      & 0.89       & 1.11                             & \textbf{1.25}                                            \\
6                                      & 0.86       & 1.17                             & \textbf{1.43}                                            \\
7                                      & 0.89       & 1.33                             & \textbf{1.44}                                            \\
8                                      & 1.02       & 1.4                              & \textbf{1.42}                                            \\
9                                      & 1.13       & 1.42                             & \textbf{1.57}                                            \\
10                                     & 1.15       & 1.51                             & \textbf{1.59}                                            \\
11                                     & 1.25       & 1.49                             & \textbf{1.58}                                            \\
12                                     & 1.28       & 1.47                             & \textbf{1.64}                                            \\
13                                     & 1.29       & 1.48                             & \textbf{1.73}                                            \\
14                                     & 1.32       & 1.54                             & \textbf{1.73}                                            \\
15                                     & 1.33       & 1.52                             & \textbf{1.73}                                            \\
16                                     & 1.37       & 1.53                             & \textbf{1.74}                                            \\
17                                     & 1.44       & 1.55                             & \textbf{1.74}                                            \\
18                                     & 1.36       & 1.56                             & \textbf{1.77}                                            \\
19                                     & 1.44       & 1.58                             & \textbf{1.78}                                            \\
20                                     & 1.42       & 1.59                             & \textbf{1.8}                                            
\end{tabular}
\caption{Naming scores of the algorithms presented in Section~\ref{sec:naming_score} applied to the $2791$ games with the tag \emph{Racing}.}
\label{table:racing}
\end{table}
\begin{table}[ht]
\centering
\begin{tabular}{cccc}
\multicolumn{1}{l}{number of clusters} & all tags   & \multicolumn{1}{l}{capital tags} & \multicolumn{1}{l}{normalised and weighted capital tags} \\
1                                      & \textbf{0} & \textbf{0}                       & \textbf{0}                                               \\
2                                      & 0.67       & \textbf{0.71}                    & 0.53                                                     \\
3                                      & 0.6        & 0.71                             & \textbf{0.77}                                            \\
4                                      & 0.7        & \textbf{1.08}                    & 1.02                                                     \\
5                                      & 0.93       & 1.12                             & \textbf{1.18}                                            \\
6                                      & 1.02       & \textbf{1.34}                    & \textbf{1.34}                                            \\
7                                      & 1.16       & 1.34                             & \textbf{1.44}                                            \\
8                                      & 1.21       & 1.41                             & \textbf{1.55}                                            \\
9                                      & 1.22       & 1.45                             & \textbf{1.6}                                             \\
10                                     & 1.35       & 1.44                             & \textbf{1.68}                                            \\
11                                     & 1.39       & 1.49                             & \textbf{1.66}                                            \\
12                                     & 1.36       & 1.51                             & \textbf{1.66}                                            \\
13                                     & 1.48       & 1.53                             & \textbf{1.72}                                            \\
14                                     & 1.51       & 1.56                             & \textbf{1.78}                                            \\
15                                     & 1.57       & 1.64                             & \textbf{1.83}                                            \\
16                                     & 1.58       & 1.65                             & \textbf{1.88}                                            \\
17                                     & 1.56       & 1.65                             & \textbf{1.88}                                            \\
18                                     & 1.53       & 1.7                              & \textbf{1.88}                                            \\
19                                     & 1.57       & 1.71                             & \textbf{1.88}                                            \\
20                                     & 1.58       & 1.71                             & \textbf{1.88}                                           
\end{tabular}
\caption{Naming scores of the algorithms presented in Section~\ref{sec:naming_score} applied to the $12185$ games with the tag \emph{Puzzle}.}
\label{table:puzzle}
\end{table}
\begin{table}[ht]
\centering
\begin{tabular}{cccc}
\multicolumn{1}{l}{number of clusters} & all tags   & \multicolumn{1}{l}{capital tags} & \multicolumn{1}{l}{normalised and weighted capital tags} \\
1                                      & \textbf{0} & \textbf{0}                       & \textbf{0}                                               \\
2                                      & 0.85       & \textbf{0.93}                    & 0.57                                                     \\
3                                      & 0.84       & 0.88                             & \textbf{1.18}                                            \\
4                                      & 0.86       & 1.29                             & \textbf{1.36}                                            \\
5                                      & 0.92       & 1.4                              & \textbf{1.42}                                            \\
6                                      & 1.05       & 1.41                             & \textbf{1.54}                                            \\
7                                      & 1.15       & \textbf{1.39}                    & 1.38                                                     \\
8                                      & 1.21       & 1.57                             & \textbf{1.6}                                             \\
9                                      & 1.2        & \textbf{1.62}                    & 1.52                                                     \\
10                                     & 1.27       & 1.64                             & \textbf{1.65}                                            \\
11                                     & 1.3        & 1.59                             & \textbf{1.7}                                             \\
12                                     & 1.33       & 1.63                             & \textbf{1.69}                                            \\
13                                     & 1.4        & 1.66                             & \textbf{1.84}                                            \\
14                                     & 1.39       & 1.66                             & \textbf{1.84}                                            \\
15                                     & 1.47       & 1.64                             & \textbf{1.88}                                            \\
16                                     & 1.53       & 1.65                             & \textbf{1.93}                                            \\
17                                     & 1.52       & 1.65                             & \textbf{1.94}                                            \\
18                                     & 1.51       & 1.68                             & \textbf{1.97}                                            \\
19                                     & 1.5        & 1.7                              & \textbf{1.99}                                            \\
20                                     & 1.51       & 1.68                             & \textbf{2}                                              
\end{tabular}
\caption{Naming scores of the algorithms presented in Section~\ref{sec:naming_score} applied to the $2961$ games with the tag \emph{Rogue-like}.}
\label{table:rogue-like}
\end{table}

Table~\ref{table:all_games} depicts the naming scores of the algorithms presented in Section~\ref{sec:naming_score} applied to the $4065$ games with at least $2000$ reviews. Likewise, Table~\ref{table:racing} does the same for the $2791$ games with the tag \emph{Racing}, Table~\ref{table:puzzle} for the $12158$ games with the tag \emph{Puzzle} and Table~\ref{table:rogue-like} for the $2961$ games with the tag \emph{Rogue-like}. 

\end{document}